\documentclass[preprint,12pt]{elsarticle}

\usepackage{amsmath}
\usepackage{amssymb} 
\newcounter{bla}

\journal{Computer Physics Communications}

\begin{document}

\begin{frontmatter}

\title{JOZSO, a  computer code  for calculating broad neutron resonances in phenomenological
nuclear potentials}

\author[a]{\'A. Baran}
\author[a]{Cs. Nosz\'aly}
\author[a,b]{T. Vertse\corref{author}}
\address[a]{University of Debrecen, Faculty of Informatics, PO Box 12, H--4010 Debrecen, Hungary}
\address[b]{Institute for Nuclear Research  Hungarian Academy of Sciences (ATOMKI),
Debrecen, PO Box 51, H--4001, Hungary}
\cortext[author] {Corresponding author.\\\textit{E-mail address:} vertse@atomki.hu}


\begin{abstract}

A renewed version of the computer code GAMOW ~\cite{[Ve82]} is given in which
the difficulties in calculating broad neutron resonances are amended. New types
of phenomenological neutron potentials with strict finite range are built in.
Landscape of the S-matrix can be generated on a given domain of the complex
wave number plane and  S-matrix poles in the domain are localized. Normalized 
Gamow wave functions and trajectories of given poles can be calculated optionally.  
\footnote{ The program name is chosen to honor   the late J\'ozsef Zim\'anyi to whom 
one of the authors (T. Vertse) is grateful for starting his carrier.}
\end{abstract}

\end{frontmatter}

{\bf PROGRAM SUMMARY}

\begin{small}
\noindent
{\em Program Title: }JOZSO                                          \\
{\em Licensing provisions: GPLv3}                                   \\
{\em Programming language:} Fortran 90                                    \\
{\em Supplementary material:} A readme file: https://github.com/czylabsonasa/jozso                           \\
{\em Nature of problem:}\\
  The program calculates the poles of the partial wave S-matrix for spherically symmetric strictly finite range complex potentials. 
A few types of potential forms are built in and option for reading in external potential form is given. Landscape of the S-matrix on a given domain 
of the complex wave number plane can be calculated. Accurate position of the poles can be determined.  Normalized 
Gamow wave functions and trajectories of given poles can be calculated optionally.   \\
{\em Solution method:}\\
  Internal and external solution satisfying boundary conditions in the origin and in the asymptotic region are generated by integrating 
the radial equation with adaptive step-size control for Runge-Kutta method. The difference of the logarithmic derivatives are calculated 
for a range of distances. The minimum of the summed modulus of the differences is searched using the Nelder-Mead algorithm. Pole 
trajectories and normalized Gamow functions can be calculated optionally.  \\ 
{\em Additional comments including Restrictions and Unusual features:}\\
  The region of interest is restricted to the lower half of the wave number plane. Pole solutions from the upper half wave number plane can be 
safely computed by using the codes GAMOW \cite{[Ve82a]} and ANTI \cite{[Ix95a]}.
   \\

\end{small}

\section{Introduction}

One of the possibility of defining resonances is by the purely outgoing
solutions of the Schroedinger equation (Siegert condition), i.e. by solutions with complex
energies at the pole of the $S$-matrix. These solutions are called Gamow
states, since Gamow was the first who introduced them in nuclear physics for the description
of $\alpha$-decay early in the last century.
Gamow states represent non-stationary states, since they describe resonant states with finite lifetime. 
The lifetime is promotional to the inverse of
the width of the resonance. The energy of the Gamow solution is
a discrete complex value which corresponds to the pole of the $S$-matrix in the complex energy sheets. 
It is more convenient to use the wave number $k$ instead
of the energy, since in single channel problem we have only one complex wave number sheet. The energy is 
proportional to the square of $k$, therefore the upper half of the $k$-plane maps to one of the energy 
sheets (called physical energy sheet), while the lower half of the $k$-plane maps to another energy sheet, called unphysical energy sheet.

 Decaying resonances lie in the fourth quadrant of the $k$-plane. For a real potential the capturing resonances
are mirror images of the decaying resonances. Both type of resonances lie on the
second energy sheet, the decaying ones are below, the capturing ones are above
the real axis, where the cut separates the first and the second energy sheets.
In the calculation of nuclear reactions we often use complex potential (optical
potential). The JOZSO program is able to calculate resonances
in complex potentials like its ancestor, the code GAMOW. The name of the new program comes from the nickname 
of the late J\'ozsef Zim\'anyi to whom one of the authors (T.V.) is greatly indebted.

One of the goal of writing a new code is to provide an efficient tool for
calculation the poles of the $S$-matrix under a whole domain of the complex $k$ plane including regions with 
large values of the complex $k$-values.  First a  landscape of  the $S(k)$-matrix on a mesh of certain region 
of the  $k$ plane is calculated, then the positions of the poles in the domain are
localized. 
After we localized a complex $k$ eigenvalue we normalize its wave function to unity, using regularization 
methods taking the contribution of the external region into account. For neutral particle the contribution 
of the external region can be calculated in closed form given in Ref.\cite{[Gy71]}. It turned out only later 
that the normalized wave function
of the anti-bound pole can be either real or imaginary in a real potential \cite{[Da12]}.
 An option for calculating trajectory of
selected poles is built into the new program. The trajectories are  complex curves in the complex wave number plane along
a certain pole moves as the potential strength slowly changes. The potential strength is a real number $\gamma$, 
which multiplies the sum of the nuclear potential terms. The $\gamma\rightarrow 0$ represents the free particle 
limit. The use of the code JOZSO was demonstrated in Refs. \cite{[Ba15]} and \cite{[Sa15]}. In a most recent work 
\cite{[Ba17]} it was found that the JOZSO program gives more accurate values for the imaginary part of the resonance for
large values of the real part of $k$.

The program deals with potential wells only, i.e.
in which the poles become bound or anti-bound for large values of the potential depth.

 We describe the mathematical formalism in details in the following sections.

\section{Formalism}
The complex energy of the Gamow solution is $E=E^R-i E^I$, where $E^R$ is the position of the
resonance and $E^I=\Gamma/2$ with the width $\Gamma$. The wave number of the Gamow resonance is also complex, 
$k=k^R+ i k^I$. Since $k^I$ is negative, therefore the complex $k$ lies on the lower half of the complex $k$-plane. 
The real part $k^R$ is positive for a decaying resonance, while a resonance with  negative real part $k^R$ corresponds 
to a capturing resonance. The energy is proportional to the square of $k$, therefore the resonant energy is on the 
second  Riemann-sheet,  
it is a discrete complex eigenvalue of the differential equation. Strictly speaking it is a generalized eigenvalue 
since $E$ is not real as an eigenvalue in normal sense should be. We assume that the phenomenological potential is 
spherically symmetric, therefore the use of the polar coordinates is convenient.
The resonant wave function satisfies the radial Schroedinger equation as follows:
\begin{equation}
\label{radsch}
u^{\prime\prime}(r,k)+\left[ k^2 -\frac{l(l+1)}{r^2}-v(r) \right] u(r,k)=0~,
\end{equation}
where prime denotes the derivative with respect to the radial distance $r$. The non-negative integer $l$ denotes 
the quantum number of the orbital angular momentum, $v(r)$ denotes the sum of the
nuclear and Coulomb potentials both having spherical symmetry.
In our case we have no Coulomb potential term.
We can rewrite the radial equation in Eq.(\ref{radsch}) into the form expressed by the so called squared local wave number:
\begin{equation}
\label{locwn}
k_l^2(r)=\left[ k^2 -\frac{l(l+1)}{r^2}-v(r)\right] ~,
\end{equation}

\begin{equation}
\label{radsch1}
u^{\prime\prime}(r,k)+k_l^2(r) u(r,k)=0~.
\end{equation}

 The solution $u(r,k)$ satisfies boundary condition (BC) at the origin $r=0$ and
at large distance $r \ge R_{as}$, beyond the range of the nuclear potential $R_{max}$.  
The energy $E$ and the potentials too are written in the same units as $k^2$ and the centrifugal term:
$\frac{l(l+1)}{r^2}$ , namely in $[fm^{-2}]$. The  factor 
$c_1=\frac{2 \mu}{\hbar^2}$ converts
from the usual MeV units to $[fm^{-2}]$ and it includes the reduced mass $\mu=\frac{m_p m_T}{(m_p+m_T)}$ of the 
projectile-target system. Therefore  
\begin{equation}
\label{ek2}
k^2=c_1 E \quad\quad v(r)=c_1 V(r)~,
\end{equation}
where $V(r)$ denotes the total nuclear potential. The nuclear potential we deal with here has a strictly finite range 
(SFR) feature, i.e it falls to zero and remains zero beyond a finite distance.

The BC at the origin requires that the solution of the radial equation is regular:
\begin{equation}
\label{regular}
u(0,k)=0~.
\end{equation}
The other BC is specified in a large distance: $R_{as}\ge R_{max}$, where
the nuclear  potential vanishes:
\begin{equation}
\label{sumpt}
V(r\ge R_{max})=0~.
\end{equation}

\subsection{Asymptotic forms of the SFR potentials and the solutions}
At $R_{as}$ i.e. at or beyond $R_{max}$ our radial equation in Eq.(\ref{radsch}) evolves to its asymptotic form without potential.
It describes free spherical waves, which satisfy the Riccati--Hankel differential equation 
\begin{equation}
\label{asimpde}
u^{\prime\prime}(r,k)+\left[ k^2 -\frac{l(l+1)}{r^2} \right] u(r,k)=0~.
\end{equation}
It is convenient to change to the dimensionless variable $\rho=kr$ in the asymptotic
differential equation. 

 For a scattering state the asymptotic BC requires that the solution $u(r,k)$
 should be a linear combination of the incoming $I_l(kr)$ and outgoing $O_l(kr)$ spherical free
 waves:
 
\begin{equation}
\label{scattbc}
u(r,k)=A [I_l(kr)-S(k)O_l(kr)]~,
\end{equation}
where $S(k)$ is the element of the scattering matrix. In this case the scattering matrix is diagonal in the angular momentum, 
therefore in the partial wave $l$ $S(k)$ is a $1\times 1$ matrix, the scattering function.
The incoming $I_l(kr)$ and outgoing $O_l(kr)$ spherical
free waves are expressed by  Ricatti-Hankel functions $H_l^{\pm}(\rho)$ . Here $H_l^+(\rho)\sim
O_l(\rho)$ and $H_l^-(\rho)\sim
I_l(\rho)$. 
 
Solutions being regular at the origin at any real or complex $k$ values
can be matched at $R_{as}$ to the combinations of $I_l(kr)$ and $O_l(kr)$ solutions of the asymptotic differential equation.

For the majority of the potential forms used in nuclear calculations the  radial equation in Eq. (\ref{radsch}) can not be solved 
analytically and the  solution $u(r,k)$ should be calculated by using numerical
solution methods with different approximations.
At $r=R_{as}$ the numerical solution should match to that of the asymptotic equation in Eq.(\ref{scattbc}).
The derivative $u^\prime(r,k)$, what we also calculate numerically should be equal to the derivative of the asymptotic equation, therefore
\begin{equation}
\label{scattbc1}
u^\prime (r,k)=A~k [{\bar I}_l(kr)-S(k){\bar O}_l(kr)]~,
\end{equation}
where ${\bar I}_l(kr)$ and ${\bar O}_l(kr)$ denote the derivatives of the incoming and outgoing Ricatti-Hankel functions  with respect to  $\rho=kr$.
The magnitude $A$ of the asymptotic solution falls out from the logarithmic derivative  if $r$ is in the asymptotic region.
We can calculate the value of $S(k)$ from the logarithmic derivative at $r=R_{as}$:
\begin{equation}
\label{lgder}
z_i(R_{as},k)=\frac{u_i^\prime (R_{as},k)}{u_i(R_{as},k)}=k\frac{{\bar I}_l(kR_{as})-S(k){\bar O}_l(kR_{as})}{I_l(kR_{as})-S(k)O_l(kR_{as})} ~.
\end{equation}
Here we denote by $u_i(r,k)$ the internal solution
being regular at $r=0$.

The value of the $S$-matrix at the wave number $k$ can be calculated from this
relation:
\begin{equation}
\label{smatr}
S(k)=\frac{k {\bar I}_l(kR_{as})-z_i(R_{as},k) I_l(kR_{as})}{k{\bar O}_l(kR_{as})-z_i(R_{as},k) O_l(kR_{as})}~.
\end{equation}

If we solve this differential equation in Eq.(\ref{radsch}) numerically, we get at $r=R_{as}$
the logarithmic derivative $z_i(R_{as},k)$ needed for calculating the
value of $S(k)$ in Eq.(\ref{smatr}). In the nuclear reaction calculations we need the
values of the $S(k)$ in each partial waves to calculate cross sections.

 For a resonance the asymptotic BC requires that the solution $u(r,k)$  should be proportional to the outgoing   
function $O_l(kr)$ and its
 derivative $u^\prime(r,k)$ should be proportional to the derivative of the outgoing function, i.e.
 
\begin{equation}
\label{reson}
u(r,k)\sim O_l(kr) \quad u^\prime(r,k)\sim k {\bar O}_l(kr)~,
\end{equation} 
or the logarithmic derivative should be
\begin{equation}
\label{resonz}
z(r,k)=k \frac {{\bar O}_l(kr)}{O_l(kr)}~.
\end{equation} 
 
This BC can be satisfied only at a discrete complex $k_o$ eigenvalue, namely at a $k$-value belonging to a  pole of $S(k)$. 
The procedure of finding the poles
of $S(k)$ will be discussed in the  section \ref{findP}.

 Since the position of the pole depends on the potential (on its strength and on its shape) we
 can calculate the pole trajectory for a potential with given radial shape by calculating the pole position $k_o$
 as a function of the potential strength. By increasing the strength the  trajectory goes to the upper 
half of the $k$-plane, where (for real potential) it becomes a bound state with real
 energy and with purely imaginary $k$. Here the solution becomes to be a square integrable real function with finite 
number of nodes $n$. We assign the $n=0$ node number to the function having zero only at $r=0$.
 The $n=1$ solution has one additional node at $r>0$, the $n=2$ solution has two additional nodes at $r>0$, etc. 
Anti-bound solutions are not square integrable since they diverge as $r\rightarrow\infty$. After normalizing them 
with appropriate regularization procedure they
 become functions being either real or purely imaginary \cite{[Da12]} and they have finite number
 of nodes.
 Resonances on the other hand have no nodes, only infinite number of zeros both in the real part and in the imaginary part of 
their wave function at different distances. The asymptotic solution $O_l(kr)$ oscillates around the
 real $r$-axis with exponentially growing amplitude as $r\rightarrow\infty$. 
 To assign a given node number to a resonance is possible only if we manage to get rid off the
 oscillations in the asymptotic region by following the pole trajectory until it becomes a bound
 state with finite node number.

\subsection{Nuclear potentials}
 A common feature of the nuclear potentials is that they differ from zero only in a finite
range $r\in [0,R_{max}]$. Their functional shapes (radial forms) can be built into this program, or might be 
read in from file (external form factor). The built in potentials might have different phenomenological forms. 
A common feature of these potentials is that the radial
equation in Eq.(\ref{radsch}) is solved  numerically.
The most frequently used potential is the Woods-Saxon (WS) potential. Most of the $V(r)$ potentials, 
including the WS  becomes zero only
at infinite distance. However, the
 nuclear potential $V(r)$ we use here, should be SFR type potential, since we match our solution at 
$R_{as}$ to that of the asymptotic differential
 equation in which no nuclear part is present\cite{[Da12]}.
The most common nuclear potential is the cut-off form of the Woods-Saxon (CWS) potential. 
The CWS can be written as a product of its strength $V_0$ and its radial shape:
\begin{equation}
\label{WSpot}
V^{\rm CWS}(r,R,a,R_{\rm max})=-V_0f_{\rm CWS}(r,R,a,R_{\rm max})~,
\end{equation}
where the radial shape is
\begin{equation}
\label{vagottWS}
f_{\rm CWS}(r,R,a,R_{\rm max})=\theta(R_{\rm max}-r)
\frac{1}{1+e^{\frac{r-R}{a}}}~,
\end{equation}
where $\theta(x)$ denotes the 
Heaviside step function, being zero for negative
and unity for non-negative arguments.
 It was
shown earlier~\cite{[Sa08]} that in the CWS potential the
positions of broad resonances considered here do depend on the value of the
cut-off radius $R_{\rm max}$~\cite{[An11],[Da12]}, therefore, the
cut-off radius is an important parameter of the CWS form in
Eq.~(\ref{vagottWS}). The two other parameters of the CWS form are
the radius $R$ and the diffuseness $a$.

The generalized WS potential (GWS) is a combination of a Woods-Saxon (WS) potential term and a 
surface term with potential
 strengths  $V_0$ and $V_1$ . The radial form of the WS term is 
\begin{equation}
\label{WSshape}
f^{\rm WS}(r,R,a)=~-~
\frac{1}{1+e^{\frac{r-R}{a}}}~,
\end{equation}
while the shape of the surface term is 
\begin{equation}
\label{SWSshape}
f^{\rm SWS}(r,R,a)=~-~
\frac{e^{\frac{r-R}{a}}}{(1+e^{\frac{r-R}{a}})^2}~.
\end{equation}
The geometrical parameters of the terms are the radius $R$ and diffuseness $a$.  Therefore the resulting GWS potential is the following:
\begin{equation}
\label{GWSshape}
V^{GWS}(r,R,a,V_0,V_1)=V_0f^{WS}(r,R,a)+V_1f^{SWS}(r,R,a)~.
\end{equation}
A big advantage of this potential is, that for $l=0$ the radial equation can
be solved  in closed analytic form \cite{Be66}. 
A SFR form of the GWS potential can be created, if we cut its radial form
at the same finite distance $R_{max}$ as we do it with the radial form of its volume
term. This SFR form $f^{CGWS}(r)$ has the property of being zero at and beyond the
distance  $R_{max}$. While the analytical solution exists only if the geometrical
parameters $R$, $a$ are the same for the volume and the surface terms,
for a numerical solution this is not needed and we can use different parameters
for the two terms.

A new type of phenomenological nuclear potential form (SV form) was introduced in Ref. \cite{[Sa08]}. 
The SV form becomes zero smoothly at a finite $R_{\rho}$ distance without an artificial cut-off and 
remains zero beyond that distance. The SV form has the attractive mathematical property that it belongs 
to the class of functions $C^\infty$, the functions of compact support as it was realized by N\'andori\cite{[Na13]}.

Here we specify the SV potential form as a product of its strength and radial shape

\begin{equation}
\label{SVpot}
V^{\rm SV}(r)=-V_0 f^{\rm SV}(r,c,\rho_0,\rho_1)~, 
\end{equation}
in which the strength $V_0\ge 0$, and the shape $f^{\rm SV}(r,c,\rho_0,\rho_1)$ is 
a linear combination of the function 
\begin{equation}
\label{distrib}
f(r,\rho)=
e^{\frac{r^2}{r^2-\rho^2}} ~ \theta(\rho-r)~,
\end{equation}
and a term containing the derivative, with respect to $r$, of the first factor, 
\begin{equation}
\label{SVder}
f^\prime(r,\rho)=-\frac{2 r \rho ^2}{(r^2-\rho^2)^2}
e^{\frac{{r^2}}{r^2-\rho^2}}~\theta(\rho-r)~. 
\end{equation}
\begin{equation}
\label{SVform}
f^{\rm SV}(r,c,\rho_0,\rho_1)=f(r,\rho_0) - c f^\prime (r,\rho_1)~.
\end{equation}
The combination parameter $c$ gives the weight of the derivative term.
For light nuclei the derivative term is not important and one can take $c=0$ \cite{[Sa14]}.

Sahu and Sahu~\cite{[SS12]} generalized the SV potential by introducing an extra parameter
$a_s$ to the derivative term of the SV form.
The formula of the SS potential \cite{[SS12]} is analogous to 
Eq.~(\ref{SVform}), where the SS form is given as:
\begin{equation}
\label{SSform}
f^{\rm SS}(r,c,\rho_0,\rho_1,a_s)=f(r,\rho_0) - c f^\prime (r,\rho_1,a_s)~,
\end{equation}
where 
\begin{equation}
\label{SSder}
f^\prime(r,\rho_1,a_s)=-\frac{2 r \rho_1 ^2}{(r^2-\rho_1^2)^2}
e^{\frac{a_s{r^2}}{r^2-\rho_1^2}}~~\theta(\rho_1-r)~,  
\end{equation}
with the extra diffuseness parameter $a_s$. 
When $a_s=1$, the SS form coincides with the SV potential~(\ref{SVpot}). 
By using $a_s \ne 1$, one naturally has more freedom in choosing the shape 
of the potential. With the usual choice  $\rho_0>\rho_1$, the range of the 
SS potential is also $\rho_0$. The SS form has the same attractive 
mathematical features as the SV potential, namely it is a $C^\infty$ function.

The external form factors are read in from the input file at specified points.
Since we need the potentials later at points not specified in advance, it is
convenient to interpolate the external forms using spline interpolations.
If we have the spline coefficients of the potentials we are able to calculate them at any point 
inside the range they are different from zero.

If we are calculating pole trajectories we can determine the spline coefficients
only once and multiple the interpolated shape with the potential strength we vary.

\subsection{Spin-orbit part of the nuclear potential}

Since the neutron has a non-zero spin ( $s=1/2$ in $\hbar$ unit) 
the  potential $v$ can be complemented by a spin-orbit term:
\begin{equation}
\label{spinorb}
V_{\rm so}^{\rm CGWS}(r,R_{\rm so},a_{\rm so},R_{\rm max})
=V_{\rm so}^{\rm CGWS}h_{\rm CWS}(r,R_{\rm so},a_{\rm so},R_{\rm max})~2(
{\bf l}\cdot {\bf s})~,
\end{equation}
with a radial form 
\begin{equation}
\label{spinorbr}
h_{\rm CGWS}(r,R,a,R_{\rm max})=-\frac{1}{r} f^\prime_{\rm CGWS}
(r,R,a,R_{\rm max})~,
\end{equation}
in which the derivative of the central potential
appears.
The spin-orbit term of the SS potential may be defined analogously:
\begin{equation}
\label{spinorbrss}
V_{\rm so}^{\rm SS}(r,c,\rho_0,\rho_1)=V_{\rm so}^{\rm SS}
h_{\rm SS}(r,c,\rho_0,\rho_1)~2({\bf l}\cdot{\bf s})~,
\end{equation}
with
\begin{align}
h_{\rm SS}(r,c,\rho_0,\rho_1)&=-\frac{1}{r}f^\prime_{\rm SS}(r,c,\rho_0,\rho_1)
=\frac{2\rho_0^2}{(r^2-\rho_0^2)^2}e^{\frac{r^2}{r^2-\rho_0^2}}\theta(\rho_0-r)\\ \nonumber
&-c\frac{2\rho_1^2}{(r^2-\rho_1^2)^4} e^{\frac{a_sr^2}{r^2-\rho_1^2}} 
\left( \frac{\rho_1^4}{r}-3r^3+2r\rho_1^2(1-a_s)\right) \theta(\rho_1-r)\label{spinorbrsv}
\end{align}
The spin-dependent factor $2({\bf l}\cdot {\bf s})$ can be calculated easily from the difference of the eigenvalues
of the total, the orbital and the spin quantum numbers:
\begin{equation}
2({\bf l}\cdot {\bf s})=j(j+1)-l(l+1)-s(s+1)~.
\end{equation}
Due to the $1/r$ factor in the spin-orbit potential, it might be singular at the origin. However, for the SS form with $c=0$ the singularity disappears.
In the general case the full potential is:
\begin{equation}
\label{fullpot}
V(r)=V_{cent}(r)+V_{\rm so}(r)~.
\end{equation}
Here the $V_{cent}(r)$ central part of the nuclear potential does not depend  on $j$,
while the spin-orbit part might depend on $j$ if the spin-orbit strength $V_{so}$ is different from zero. In certain cases the central
potential might depend on the orbital angular momentum $l$. The pair of the
$l,j$ quantum numbers define a given partial wave of the scattering problem.
In a given partial wave the nuclear potential might depend on $l$ and $j$, i.e. 
$V^{l,j}(r)$ and the local wave number also might depend on $j$ as well $k_{l,j}^2(r)$. The radial wave function $u(r,k)$ depends on the potential used
in the given partial wave, but this dependence is not shown explicitly in the notation for the
sake of simplicity. The value of the $S$-matrix in the given
partial wave depends on $j$ too, if the spin-orbit part of the potential is different from zero.
When we calculate pole trajectory, the potential in Eq. (\ref{fullpot}) is multiplied by the strength $\gamma$ and the value of  $\gamma$ is changing along the trajectory.

\subsection{Normalization of the wave function}
The non-normalized resonance solution we have already in $u_i(r,k_o)$ when the pole finding
procedure has converged to the pole at $k_o$. So the
logarithmic derivative of $u_i(r,k_o)$ is equal to the logarithmic derivative
of the asymptotic solution $H_l^+(k_o r)$. In order to continue the internal solution
smoothly into the asymptotic region we should multiply with it with the factor:
$F_c=\frac{u_i(R_{as},k_o)}{H_l^+(k_o R_{as})}$.

The normalization of this radial wave function of the Gamow resonance can be done
most conveniently by using the method given in Ref.\cite{[Gy71]}. The same method was implemented in the program GAMOW \cite{[Ve82]} and also in the
program ANTI \cite{[Ix95]}.
The contribution to the square of the norm from the internal region can be calculated by
quadrature as: 
\begin{equation}
\label{normint}
N_i^2=\int_0^{R_{as}} u_i(r,k)^2 dr~.
\end{equation}
This should be complemented by the contribution of the external region given
in closed form \cite{[Gy71]} as
\begin{align}
N_e^2&=F_c^2~\int_{k_oR_{as}}^\infty H_l^+(\rho)^2d\rho\\ \nonumber
&=\frac{-R_{as} F_c^2}{2}[H_l^+(k_oR_{as})^2+H_{l+1}^+(k_oR_{as})^2-\frac{2l+1}{k_oR_{as}}H_l^+(k_oR_{as})H_{l+1}^+(k_oR_{as})].\label{nnorm}
\end{align}


Now the full squared norm is
as follows
\begin{equation}
\label{norm}
N^2=N_e^2+N_i^2~,
\end{equation}
while the normalized radial wave function is:
\begin{equation}
\label{norm}
u(r,k)=\frac{1}{N}  u_i(r,k)~.
\end{equation}

\subsection{Numerical integration of the radial equation}

 For the numerical integration of the radial equation we use 
the adaptive step-size control for Runge-Kutta-Fehlberg with Cash-Karp parameters as in  \cite{[numrec]} (SUBROUTINE ODEINT with RKQC).

\subsection{Finding the poles of $S(k)$.}\label{findP}
We have more chance for finding the pole if we can start the iterations with good $k$ starting
value. In this case the calculation is very efficient, since we are solving a set of
initial value problems instead of eigenvalue problem as follows.

In the  eigenvalue problem a possibility for finding the pole of $S$ is that we calculate the external solution with 
outgoing wave starting condition at  $R_{as}$ and
from these initial values at  $R_{as}$ we propagate the numerical solution inward and calculate the external solution $u_e(r_j,k)$ 
in the mesh-points $r_j$ needed to compare it to the internal solution.

From the BC at the origin we can start an internal solution of the radial equation in Eq.(\ref{radsch}) with the starting values:
\begin{equation}
\label{internalbc}
u_i(0,k)=0 \quad\quad u_i^\prime(0,k)=1~.
\end{equation}
Both BC (in Eqs. (\ref{regular}) and  (\ref{resonz})) can be satisfied simultaneously only at discrete complex $k$ eigenvalues 
belonging to a  poles of $S(k)$. Here the complex $k$ eigenvalue  is fixed by the zeros of the the difference of the logarithmic 
derivatives of the internal and the external solutions:
\begin{equation}
\label{logder}
G(k,r)=z_i(r,k)-z_e(r,k)~, 
\end{equation}
where 
\[
z_i=\frac {u_i'(r,k)}{u_i(r,k)}\quad \text{and} \quad z_e=\frac {u_e'(r,k)}{u_e(r,k)}. 
\]
The computer programs GAMOW\cite{[Ve82]}, and ANTI\cite{[Ix95]}  find the zeros of $G(k,r)$, at certain $R_m$ matching radius $0<R_m<R_{as}$.
 For a broad resonance the proper choice of this $R_m$ is difficult. The zero  is searched by Newton iterations, and the iteration
process often converges poorly or fails. Therefore in the JOZSO program we extend the comparison of the logarithmic derivatives of $z_i$  
and $z_e$ to a wider region of $r$ and
we search for the absolute minimum of the following function:
\begin{equation}
\label{minima}
F(x_1,x_2)=log \left[ \sum_{j=i_1}^{i_2}  \left\vert G(k,r_j)\right\vert \right] ~.
\end{equation}
where $r_j$ is a mesh with equidistant mesh-points of the interval $r=[r_1,r_2]$, with step-length $h$.
 Here the two real variables in the argument of $F$ are the real and the imaginary parts of the complex  
wave number $k$ (or vice versa). The logarithmic 
derivatives are complex, hence in order to have a real valued function we take the modulus of their differences. Therefore the function 
$F$ is a real valued function. Although we calculate the logarithmic derivatives of the internal and the external 
solutions in a wide range of $r$, in the sum in Eq. (\ref{minima}) we
 include only a subinterval in which the nuclear potential falls to the certain
 fraction to their value close to the origin. We take the value of the potential close to the origin, say e.g. $V_{or}=V_N(h)$
 and select the index $i_1$ where $V(i_1*h=r_1)\approx V_{or}/10$. The higher
 value of the subscript is taken as $V(i_2*h=r_2)\approx V_{or}/1000$.

 The task is to find the absolute minimum of the real function $F$. 
 If the two solutions could be calculated without any errors than the value of the absolute minimum of the sum in the argument of 
the logarithm were zero. 
But in the numerical solution rounding errors accumulate as we proceed from the starting points, where the initial 
conditions are specified, therefore this value is somewhat larger than zero. 
 The minimum of the function $ F(x_1,x_2)$ in Eq.(\ref{minima})
should have a negative value with a large modulus.
 
 To find the minimum of the function  $F(x_1,x_2)$ we use the Nelder--Mead method.
 We are searching for
 a minimum of the function starting from some first guess  $k_0$. The two-variable  
functions $ F(x_1,x_2)$ in Eq.(\ref{minima}) have special shapes, therefore it is useful to generate 
a landscape of the $ F(x_1,x_2)$ over a grid of a region of the complex
 $k$ domain. The mesh in $x_1$ and  $x_2$ should be fine enough to localize the poles
 of $S(k)$ and at the same time the number of mesh-points should  remains within a reasonable limit. 
   The landscape helps us to supply reasonably good starting values for finding
   the absolute minima of the function $ F(x_1,x_2)$ and give the position of
   the pole with higher accuracy. Pole positions calculated agree with those given in Ref.\cite{[Ix95]} in 3-4 decimal digits.

\subsection{Starting integration from the origin}

The centrifugal term for $l>0$ has a singularity in Eq.(\ref{locwn}), therefore it might be convenient from numerical point of view to use an expansion method for the
regular solution close to the origin.
Here (at the first $nbo$ points) we can expand the internal solution $u_i(r,k)$ into
powers of $r$ and we search for the solution for $r<<1$ in a form:
\begin{equation}
\label{powexp}
u_i(r,k)=\sum_{j=0}^{50} a_j r^{l+1+j}~.
\end{equation}
The full potential in Eq.(\ref{sumpt}) can be split into two terms, the one which is
singular at the origin ($v_{sg}(r)$)  and the rest which is not singular ($v_{ns}(r)$). For a neutron the only
term which might be singular at $r=0$ is the spin-orbit potential $v_{so}(r)$ with radial form in Eq.(\ref{spinorbr}). The spin-orbit potential is non-zero
only for $l>0$ and for non-zero spin-orbit strength, in this case it can be approximated as $b/r$.
For the SS potential the second term (Eq.(\ref{SSder})) is proportional to $r$, 
therefore $b=0$.    
 Let us approximate the non-singular potential with a parabola and write the full potential for $r<<1$ as
\begin{equation}
\label{fpot}
v(r)\approx b/r+v_0+v_1 r+v_2 r^2~.
\end{equation} 
With this the local wave number in Eq.(\ref{locwn}) can be approximated as 
\begin{equation}
\label{locwnap}
k_{l,j}^2(r)=a-v_1 r-v_2 r^2 -\frac{l(l+1)}{r^2}-\frac{b}{r}~.
\end{equation}
The coefficients of the expansion in Eq.(\ref{powexp}) can be calculated as:
$$ a_0=1 \quad a_1=\frac{a_0b}{2l+2}\quad a_2=\frac{a_1b-a_0a}{4l+6}\quad a_3=\frac{a_2b-a_1a+a_0v_1}{6l+12}$$ and for $i>3$ the following recurrence relation holds
\begin{equation}
\label{reca}
a_i=\frac{1}{i(2l+i+1)}[a_{i-1}b-a_{i-2}a+a_{i-3}v_1+a_{i-4}v_2]~.
\end{equation}
The derivation of the form in Eq.(\ref{powexp}) is straightforward, and it gives $u^\prime(r,k)$ at the first equidistant mesh-points: $r_j=j*h$, for $j=1,\dots,nbo$, 
and the logarithmic derivatives will be 
\begin{equation}
\label{powexp1}
z_i(r,k)=\frac{\sum_{j=0} a_j (l+j+1) r^{l+j}}{\sum_{j=0} a_j r^{l+1+j}}~.
\end{equation}

Having the solution
and its derivative at $r_{nbo}=nbo*h$ we can proceed for $r>r_{nbo}$ and calculate the solution using the integration routine DIFFSOLVE   with the required 
accuracy $eps$. This way we propagate the internal solution $u_i(r,k)$ or the logarithmic derivative function $z_i(r,k)$ from this $r_{nbo}$ 
point outward, until we reach the asymptotic region at
$r=R_{as}$.

\subsection{Calculation of the asymptotic solutions}

For neutron the asymptotic solution  goes to the differential equation of the Ricatti-Hankel functions $H_l^{\pm}(\rho)$ . Here $H_l^+(\rho)\sim
O_l(\rho)$ and $H_l^-(\rho)\sim
I_l(\rho)$ .
 \begin{equation}
\label{freede}
\frac{d^2 w(\rho)}{d\rho^2}+[1 -\frac{l(l+1)}{\rho^2}]w(\rho)=0~.
\end{equation}
The Ricatti-Hankel functions can be calculated easily
by using the three terms recurrence relation:
\begin{equation}
\label{rechan}
H_{l+1}^\pm(\rho)=\frac{2l+1}{\rho}H_l^\pm(\rho)-H_{l-1}^\pm(\rho)~.
\end{equation}
To start the recurrence we use the known form of the $l=0,1$ functions:
\begin{equation}
\label{starth}
H_{0}^\pm(\rho)=e^{\pm i\rho} \quad\quad H_{1}^\pm(\rho)=(\frac{1}{\rho}-i)e^{\pm i\rho} ~.
\end{equation}

For $l=0$ the derivative of the
 Ricatti-Hankel function is simply $ {\bar H}_{0}^\pm(\rho)= \pm i  H_{0}^\pm(\rho)$.
 For $l>0$ the derivatives   can be calculated from the relation:
\begin{equation}
\label{derhan}
{\bar H}_{l}^\pm(\rho)=H_{l-1}^\pm(\rho)-\frac{l}{\rho}H_{l}^\pm(\rho)~.
\end{equation}

For a Gamow resonance the external BC fixes the external solution of the radial equation as:
\begin{align}
\label{externalbc}
u_e(r=R_{as},k)&=O_l(kR_{as})=H_l^+(kR_{as}), \\ u_e^\prime(r=R_{as},k)&=k {\bar O}_l(kR_{as})=k {\bar H}_l^+(kR_{as})~,\nonumber
\end{align}
with the actual value of the complex wave number $k$. We can propagate the numerical solution from this point inward and calculate the external solution $u_e(R_{as},k)$.

\subsection{Trajectory calculation}

For the calculation of the trajectory of a given pole, we can introduce a real
parameter $\gamma$ for the nuclear potential and define the nuclear potential
as a product of the full nuclear potential and this strength $\gamma$.
Then we consider the pole position as a function of the real variable $\gamma$.

We start from a strength $\gamma=1$ and increase or decrease the strength in
certain $\delta \gamma$ steps. The number of steps in $\gamma$ is limited by an input data of
the program.
The sign of the step  $\delta \gamma$ defines if we go up or down along the trajectory. 
A positive step $\delta \gamma>0$  increases the $\gamma$ and we use deeper and deeper potentials.
The negative value has the opposite effect and we proceed downward along the
trajectory toward the starting (small) value of the potential. The move of the
poles along their trajectories was studied recently in Ref. \cite{[Sa15]}.

If we have the pole wave number of the resonance $k_0$ at a potential strength
$\gamma$ we can start to calculate pole trajectory $k_0(\gamma)$ starting from this point. 
We modify the potential strength with a small $\delta \gamma$ value and
calculate the pole with this modified value to get $k_0(\gamma+\delta\gamma)$.
If the change $\delta\gamma$ is small then the change of the $\delta k_0=k_0(\gamma+\delta\gamma)-k_0(\gamma)$ 
is also small and the convergence to the pole of the changed potential will be
fast and reliable.

We can calculate the pole trajectories from their starting points where $\gamma$
is very small and we can proceed by increasing their strength. The advantage of
using this tactics is that the starting values of the trajectories show regularity for the strictly finite range potentials as has been discussed in
Ref.\cite{[Sa14]}. The shape of the trajectory naturally depends on the radial
shape of the potential and also on the partial wave $l,j$ and the
sequence number $m$ of the pole.
The index $m$ increases as the $k^R$ increases, the $m=1$ is the pole being closest to the imaginary $k$-axis. The index $m$ is different from the 
 node number $n$ what
we can determine only when we continue the trajectory to the upper half of the
$k$-plane and we have a bound state pole with definite number of nodes. The larger is the index $m$ the farther is the starting  point $k_s$ 
from the imaginary $k$-axis, and we have to follow the pole along a long way until
the resonance becomes a bound state in a deep potential.

We can follow another tactics and start the trajectory from the bound state in a deep potential, where we can count its node easily. Then we make the potential shallower 
by taking $\delta \gamma <0$ and proceed 
along the trajectory until we cross the origin $k=0$ and go to the lower
half of the $k$-plane. For $l>0$ where we have a centrifugal barrier the bound
state pole goes to a resonance, while for $l=0$ (no barrier) the bound state first becomes to be anti-bound state and might become resonance when the pole
departs from the imaginary $k$-axis. This situation is studied in Ref.\cite{[Da12]} extensively.

\subsection{The use of external potential form}
Sometimes it is useful to use potential form calculated separately, by using
different theoretical models.
In this case the geometrical form can be given in a file {\it extformf.dat} at
equidistant $h$ steps in $r$.
After reading in the external radial form, we use spline interpolation based on
the read in knots in order to be able to calculate the external form at any
distance needed by the numerical integration routine diffsolve.
The complex strength of the potential is given in P(18)+i P(19) as with other
potential form.
A limitation of the present program is that there is no spin-orbit term in the external form. 
If one wants to include spin-orbit potential one has to add it
to the external potential form to be read in.

\section{Program structure}
Input data are read from the file cgws.config or sahu.config or ext.config. 
There are  three possible types of potentials, the type of the potential can be controlled by a 
command line parameter. \\
jozso -1 or jozso -cgws uses CGWS potential,\\
jozso -2 or jozso -sahu uses SS potential,\\
jozso -4 or jozso -ext uses an external potential,\\
the default potential type is the CGWS. \\ 
The program can be used for calculation a single  pole (starting from an initial guess), or 
for calculation all the poles on a given domain, or for calculation a pole trajectory 
starting from a given pole. These options are controlled by the MODE parameter. \\
The parameters read in are stored in array P(1),...,P(45).\\
Meaning of the parameters similar to that in the program GAMOW, but some of the
parameters are not used or have different role.\\
Here we list the meaning of the input parameters:\\
P(1)-P(7) have the same meaning as in GAMOW, namely:\\
P(1)=AT mass of the target,\\
P(2)=ZT, charge number of the target (not used),\\
P(3)=AP mass of the projectile,\\
P(4)=ZP charge number of the projectile (not used),\\
P(5)=LP orbital angular momentum of the projectile,\\
P(6)=JP total angular momentum of the projectile,\\
P(7)=SP spin of the projectile,\\

\vskip .3cm
\noindent
P(8)=NBO number of points where series expansion is used.\\
Parameters P(9)-P(10) are the initial value of the wave number for MODE=0 and 
MODE=2. They are not used for MODE=1. \\
P(9)=KR is the real part of the starting value of the complex $k$.\\
P(10)=KI is the imaginary part of the starting value of the complex $k$.\\

\vskip .3cm
\noindent 
Parameters P(11)-P(16) are used for MODE=1 only. \\
P(11)=MESHKR number of mesh-points for the real part of $k$\\
P(12)=MESHKI number of mesh-points for the imaginary part of $k$\\
P(13)=DOMAINKR1 lower limit for the real part of $k$\\
P(14)=DOMAINKI1 upper limit for the imaginary part of $k$\\
P(15)=DOMAINKR2 upper limit for the real part of $k$\\
P(16)=DOMAINKI2 lower limit for the imaginary part of $k$\\

\vskip .3cm
\noindent The parameters P(17)-P(31) are the potential parameters. \\
P(17)=NX gives the type of the nuclear potential.\\ 
The value of NX is given not in the input file, it is determined from the name of 
the input file.\\ 
 NX=1 denotes the CGWS potential\\
 NX=2 denotes the SS.\\
 NX=4 denotes the external potential to be 
read in from the file extformf.dat.\\ 
The  GWS and SS  nuclear potentials are the sum of the central and spin-orbit potential terms. 
The sums are multiplied by the strength $\gamma$, when pole trajectory is calculated.

\vskip .3cm
\noindent Parameters of the central potential if NX=1:\\
P(18)=VR is the real part of the volume term of the CGWS potential,\\
P(19)=VI is the imaginary part of the volume term of the CGWS potential,\\
P(20)=R0 is the radius parameter of the volume term, the radius is R0$\cdot $AT$^{1/3}$\\
P(21)=A is the diffuseness of the volume term,\\
P(22)=VR1 is the real part of the surface term,\\
P(23)=VI1 is the imaginary part of the surface term,\\
P(24)=R01  is the radius parameter of the surface term, the radius is R01$\cdot $AT$^{1/3}$\\
P(25)=A1  is the diffuseness of the surface term,\\
P(26)=GAMMA0 the starting value of the $\gamma $ parameter used for calculation of  pole trajectories.\\

\vskip .3cm
\noindent Parameters of the central potential if NX=2:\\
P(18)=VR is the real part of the volume term of the SS potential,\\
P(19)=VI is the imaginary part of the volume term of the SS potential,\\
P(20)=RHO0 is the range $\rho_0$ of the first term of the
SS potential, it must be equal to $R_{max}$ in P(38).\\
P(21)=RHO1 is the range $\rho_1$ of the second term of the
SS potential, it has to be smaller than $\rho_0$,\\
P(22)=CS is the relative weigth  of the  second term of the
SS potential wrt. the first term, $c$ in Eqs. \eqref{SVform} and \eqref{SSform}.\\
P(23)=AS is the diffuseness parameter $a_s$ in the second term of the
SS potential. The $a_s=1$ means SV potential,\\
P(26)=GAMMA0 the starting value of the $\gamma $ parameter used for calculation of the pole trajectories.\\

\vskip .3cm
\noindent P(27)=NXSO 

\vskip .3cm
\noindent If NXSO=3 then the 
parameters of the spin-orbit potential for NX=1 are the following:\\
P(28)=VSOR is the real part of the spin-orbit strength,\\
P(29)=VSOI is the imaginary part of the spin-orbit strength,\\
P(30)=R0SO is the radius parameter of the spin-orbit potential for CGWS, the radius is R0SO$\cdot $AT$^{1/3}$\\
P(31)=ASO is the diffuseness parameter  of the spin-orbit potential for CGWS.\\

\vskip .3cm
\noindent If NXSO=3 then the 
parameters of the spin-orbit potential for NX=2 are the following:\\
P(28)=VSOR is the real part of the spin-orbit strength,\\
P(29)=VSOI is the imaginary part of the spin-orbit strength,\\

\vskip .3cm
\noindent For NX=4 the there is only the complex strength in P(19)+iP(19) and the real strength is: \\
P(26)=GAMMA0 (the starting value of the $\gamma $ parameter used for calculation of the pole trajectories).\\

\vskip .3cm
\noindent P(37)=H step size in $r$, where the radial wave function is calculated.\\
P(38)=RMAX The nuclear potential vanishes beyond this distance. For SS potential it should be equal to RHO0.\\
P(39)=RAS it is the distance $R_{as}$ where the asymptotic solution is matched to the numerical one. RAS should be 
greater than RMAX. \\
P(40)=DGAMMA is the step size of the change of the potential strength $\gamma$.\\
P(41)=NGAMMA is the maximal number of changes of the potential strength $\gamma$.\\
P(42)=C1 (conversion factor between the energy and the $k^2$) \\
P(43)=THRESHOLD threshold value for $-F(k)$. Only $-F>$P(43) peaks are selected as poles.\\
P(44)=MODE \\
For MODE=0 the program searches for a single pole starting from the initial value given by KR and KI. \\
For MODE=1 the program calculates the values of $F$ at the mesh points of the rectangle given by DOMAINKR1, DOMAINKI1, DOMAINKR2, 
DOMAINKI2, MESHKR, MESHKI. The mesh points and the function values are written to the file map.dat. The local minima of the function 
$F$ are determined and written to the file peaks.dat. These points are used as starting points for the pole calculation. The poles calculated are 
written to the file poles.dat.\\
 For MODE=2
the program calculates pole trajectory by changing the potential strength by DGAMMA. The change is performed maximum NGAMMA times\\
P(45)=WF calculates the normalized wave function if wf=1 and writes it to the file wf.dat. \\

Output files in different options.\\
The general output file is jozso.res. 
Special output files for different options:\\

For mode=0 the wave number eigen-value $k$ is written to the general output file
and on the screen as well.\\

For mode=1
the function values of the $F$ are written to the file map.dat in each mesh points in $k$.
After that the program searches the peaks for which $-F>$THRESHOLD  and writes them to file peaks.dat. 
Then it starts searching the minima of the function $F(k)$ with starting values of each approximate value. 
The values of the positions of the minima (poles) are written to file: poles.dat.\\

For mode=2 the pole trajectory is written to the file ktraj.dat (only the $k$ values) and also to the 
pltraj.dat in which the potential strength $\gamma$
is also written in the first column.

\section*{Acknowledgment}
Authors are grateful to I. Horny\'ak and P. Salamon for valuable discussions.
This work was partially supported by the OTKA Grant No. K112962.

\bibliographystyle{elsarticle-num}

\end{document}